\documentstyle[12pt,epsfig]{article}
\begin{document}
{\Large\bf B-meson hadroproduction cross sections and up-to-date models }
\\[2mm]

{\it   O.I.Piskounova}\\[2mm]

{\small P.N.Lebedev Physical Institute, Moscow}

\vspace{1cm}

\hspace{7cm}{...$"\tau$o $\tau\eta\varsigma$ $\Pi\upsilon\vartheta\iota
\alpha\varsigma$ $\gamma\rho\alpha\mu\mu\alpha$
$\varphi\rho\alpha\varsigma\varepsilon\iota$ $\tau\alpha
\nu\upsilon\nu$."}

\hspace{10cm}{$\Pi\lambda\alpha\tau\omega\nu$, No$\mu$o$\iota$ 923 $\alpha$
 5.\footnote{Plato, Laws, 923 a 5.}}

\vspace{1cm}
\large

\begin{abstract}

The comparison of B-meson production cross sections as the results of
PYTHIA code and Quark Gluon String Model (QGSM) is carried out for
energies of proton colliders: $Sp\bar(p)S$, Tevatron and LHC. Model
predictions are based on the idea of supercritical Pomeron exchanges with
the phenomenological intercept $\Delta_P$(0)=0,3 for heavy quark production.
Transverse momentum spectra of B-mesons are also compared. It is shown
that the cross sections calculated with PYTHIA using CTEQ structure functions
are in a contradiction with the asymptotical estimation of B$\bar{B}$
production cross sections in QGSM. Asymmetries between the spectra of 
$B^0$ and $\bar{B0}$ mesons are also contradicting. The reasons of the 
difference are discussed.

\end{abstract}

\newpage

\section{Introduction}

We can't say that the complete knowledge on beauty quark pair production
is obtained now, since the data we have on $b\bar{b}$ production cross sections
are not yet sufficient. In order to monitor the model ideas on the phenomenon
it seems useful to revise once in a while the  collected data. Recently the
results of several experiments \cite{UA1,CDF,D0} carried out at two energies of
colliding protons, 630 GeV and 1.8 TeV, are represented in the
literature.

In this article the two of the models are compared: on the one hand - the
phenomenological Quark-Gluon String Model \cite{4}, based on idea of
hadronic amplitude duality and on the theory of supercritical Pomeron
, on the other hand - the wide-spread Monte Carlo code  PYTHIA
 \cite{5}, which includes the results of QCD perturbative
diagram calculations.

The production cross sections arising with energy is a fact which was widely
discussed in recent studies \cite{crossections,previous}. The theory of supercritical 
Pomeron  postulates the rising as $s^{\Delta_P(0)}$, where $\Delta_P(0)=\alpha_P(0)-1$,
$\alpha_P(0)$ is the intercept of Regge trajectory of Pomeron.

The energy behavior of the production cross section in QCD perturbative 
approach are provided by the choise of gluon structure functions of 
interacting hadrons. Most 
of those functions, which are accepted now for MC simulations of high energy 
collisions, are built to approximate the recent data from HERA measured up to
$x=10^{-4}$. It should be noticed, that all known gluon structure functions 
which satisfied this recent data, can be taken for the modelling of $b\bar{b}$
production at LHC, because the value $10^{-4}$ is the very region of x 
attributed for B meson production at 14 TeV due to the following estimation:
$2m_B/(14 TeV)\sim 10^{-4}$. One of those appropriate gluon distributions is 
CTEQ structure function, which is involved into PYTHIA code as default 
distribution.

\section{Parameters defining the B-meson production cross sections in QGSM.}

 The major parameter of QGSM which defines the cross section dependence
on energy is  $\Delta_P(0)_{eff}$, which have to be less than the same
value for one Pomeron exchange, because multipomeron diagrams or branches
should be taken into account in the calculations. This parameter depends
on the mean value of transverse momenta transmitted in the process. Therefore
the energy dependences of different mass particle production cross section
must be led by different $\Delta_P(0)_{eff}$.

\begin{figure}[thb]
\centerline{\epsfig{figure=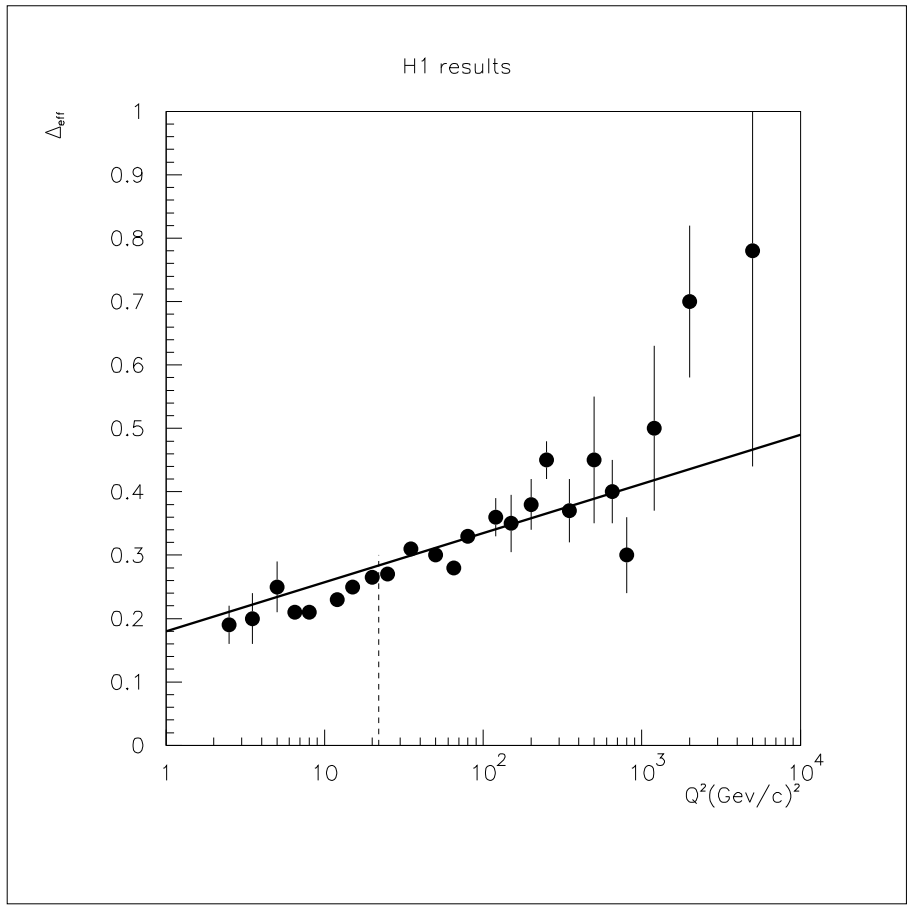,height=12cm,width=12cm}}
\caption{\label{h1}The $\Delta(Q^2)$ dependence obtained in H1 
\protect\cite{h1}
and one Pomeron exchange approximation (the solid line).}
\end{figure}

The $\Delta_P(0)(Q^2)$-dependence can be illustrated well with the data obtained
in H1 experiment at HERA \cite{h1}. The Pomeron exchange plays an important
role in electron-proton collisions too. The multyhadron production takes place
in this process due to the Pomeron exchange between photon and beam proton.
The $\Delta(Q^2)$ data are shown in Fig. \label{h1}, where each point was extracted
from measured $F_2(x,Q^2)$ function by the approximation with the simple
dependence: $F_2 \sim x^{-\Delta(Q^2)}$.

It should be noticed that the theoretical curve was calculated \cite{h1}
under the assumption of one Pomeron exchange which doesn't include the
branches as against the proton-proton interaction. The QGSM scheme for heavy
meson production have to be similar to the one Pomeron exchange pattern
due to the energy conservation reasons. So we can take for the $\Delta_{eff}$
the value 0,3 corresponding to the H1 data approximation at $Q^2=(2m_B)^2$. It's 
worth beeng stressed here, that this
value differs from $\Delta_{eff}$=0.12 which was chosen for light quark meson
production in early papers \cite{previous}.

Another Regge trajectory parameter is important for model description of
inclusive cross sections of B meson pair production, it's
$\alpha_{\Upsilon}(0)$, the intercept of $b\bar{b}$-trajectory.
It provides the increasing cross section
at the energy region close after the B pair production threshould.
There are various opinions about the
value of this parameter. From the point of view of QGSM it may  vary in the
range of -16$\div$0 \cite{previous}. The other athors prefere to take
$\alpha_\Upsilon(0)=-9$ \cite{likhoded}.

The value $\alpha_{\Upsilon}(0)=-16$ will be taken here for to estimate the
upper limit of growing cross section when it increases rapidly after the
threshould. The parameter discussed above exists in the functions
of fragmentation of qurk-gluon strings into each sort of B-mesons. Those
functions are written in QGSM according to the rules fulfilled by the
Regge asympthotics \cite{rules}.

For example, the function for d quark string fragmentation into $B^+$
contains the following factors:

\begin{displaymath}
{\cal
D}_{d}^{B^{+}}(z)=\frac{a^B_0}{z}(1-z)^{-\alpha_{\Upsilon}(0)\alpha_{\Upsilon}(0)
+\lambda}(1+a_{1}^{B}z^{2}),
\end{displaymath}

where $a^B_0$ is the density parameter for fragmentation of quark-gluon
string into B-mesons.
The $a_1^D$  is  the parameter  of  string fragmentation asymmetry
introduced in \cite{previous} to provide a transition between
probabilities of the B production at z$\rightarrow$0 and
z$\rightarrow$1.The value $a_1^B$ can be of the order 10 and actually doesn't
impact on the value of  B production cross section at  energies  higher than
1.8 TeV.

The calculations of $p_{\perp}$ spectra of produced  hadrons are also
available  in the framework of  QGSM, as it had been done already in
\cite{pt} for $\pi$-mesons. The spectra can be described up to the
momenta of order few GeV/c in  this
substantially  nonperturbative model . The distributions were of 
exponential character at low $p_{\perp}$ in this approach.   Therefore  
the transverse momenta
distributions for heavy flavor particles was not elaborated in this model.

\section{PYTHIA machinery}

The version PYTHIA 5.7 was taken to calculate  the spectra of B-mesons at
three energies of colliding protons:  630 GeV, 1,8 TeV and 14 TeV.  The CTEQ
gluon structure function \cite{CTEQ} are used  in  this  version to discribe  the
increasing cross sections. On the one hand the process of production of such
heavy quarks as b is good enough for beeing described by perturbative
QCD diagram with gluon-gluon fusion. On the other hand,
more and more low x gluons are involved into this process at energy rising
and cross section becomes dependent on the accuracy of gluon structure
function measured at low x.

As it was  mentioned in Introduction, we have precise data on $F_2$ due to
HERA experiments up to $x\sim10^{-4}$, what is enough for the calculation
of $b\bar{b}$ production at LHC energies. Such a way CTEQ structure 
functions have to provide 
the right description of increasing cross sections of $b\bar{b}$ pair production.

However, b quarks can be obtained not only in gluon fusion process. Two
additional ways exist to produce $b\bar{b}$ pair, they are: gluon 
splitting ${\it
gg \rightarrow gg}$, where gluons gives $b\bar{b}$ pair in the next-
to-leading order of corrections, and heavy flavor exitation 
${\it Q_ig \rightarrow Q_ig}$. In PYTHIA this subprocesses are taken into account with
massless matrix elements. It is a problem how to sum the resulting
distributions from such different deposits.

\begin{figure}[thb]
\centerline{\epsfig{figure=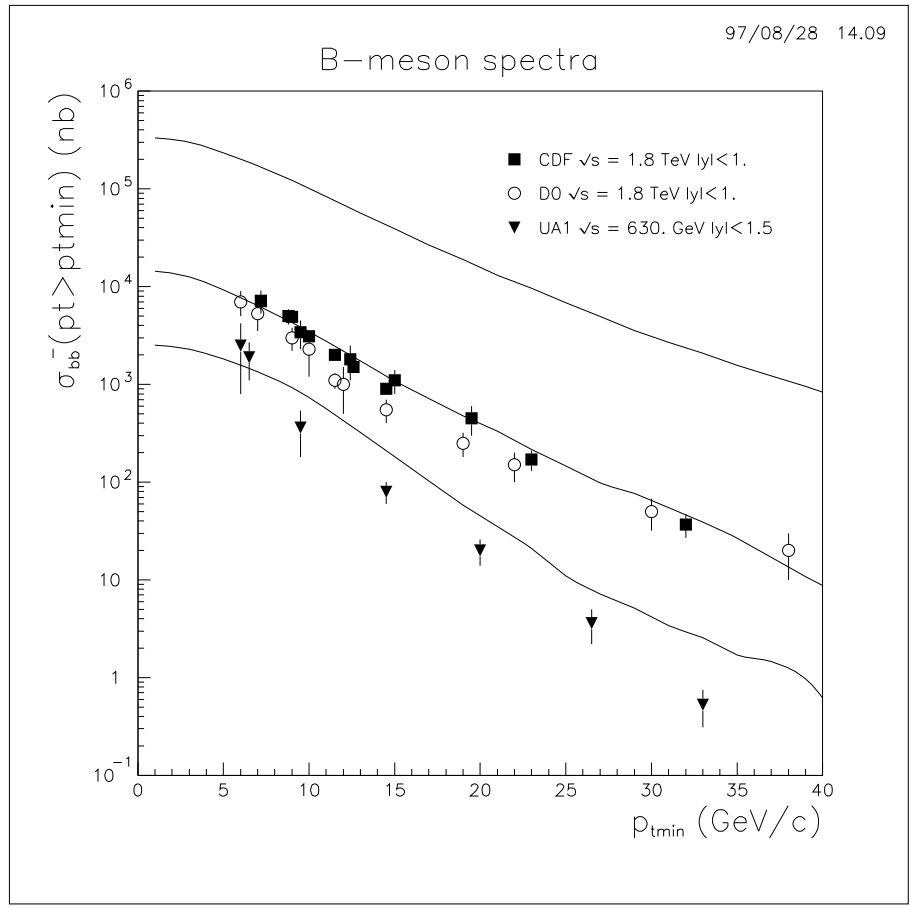,height=12cm,width=12cm}}
\caption{\label{bptintg}Transverse momenta distributions of B-mesons 
fitted with PYTHIA.}
\end{figure}

It  makes the $p_{\perp}$-spectra at 1.8 Tev comparable with  the data
obtained  in  CDF experiment ( see  Fig.\label{bptintg}). At the same time there  is not good
description  of UA1  data . It looks  like  the $p_{\perp}$ spectra were increased
with additional fractions  only by a factor   and  there is not any
difference between the patterns of spectra for different subprocesses.
It  leads to  rather flat form of transverse  momentum distributions in
 PYTHIA  and to small total cross  section  of  B production at various  energies.

\section{Comparison of cross section energy dependences}

The resulting energy dependences of production cross  section are shown 
on Fig.\label{sigmbb}
for PYTHIA program and for  QGS  model as well.

\begin{figure}[thb]
\centerline{\epsfig{figure=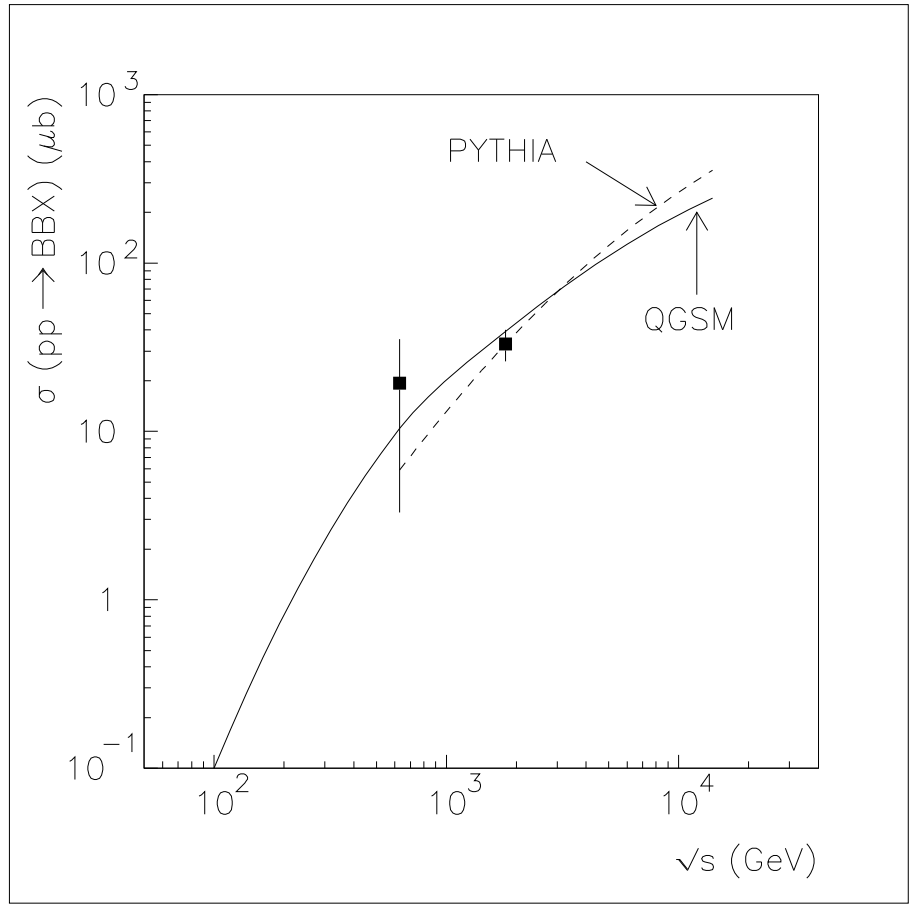,height=12cm,width=12cm}}
\caption{\label{sigmbb}Energy  dependences of B-meson cross  section.}
\end{figure}

As it was mentioned above the QGSM
curve is highest as it is possible in this model after the normalisation to 
the CDF cross section. But latter was obtained due to PYTHIA (see Fig.
\label{sigmbb}).Thus the point where both  curves are crossing is rather 
conventional and depends on the form of $p_{\perp}$-spectra  at low 
transverse momenta accepted in PYTHIA.

\section{PYTHIA and QGSM predictions for the asymmetry between 
$B^0/\bar{B^0}$ spectra }

It would be interesting to consider the leading effect in the spectra of
B-mesons at various energies. The valuable asymmetry between $B^0$- and 
$\bar{B^0}$-meson spectra at LHC energy will be important for CP violation
measurements.The recent prediction of the y dependence of such
leading/nonleading asymmetry \cite{norrbin} provided with PYTHIA simulations 
gives zero value of A(y) in wide range of y at the central region 
(see Fig. \label{asym}).

\begin{figure}[thb]
\centerline{\epsfig{figure=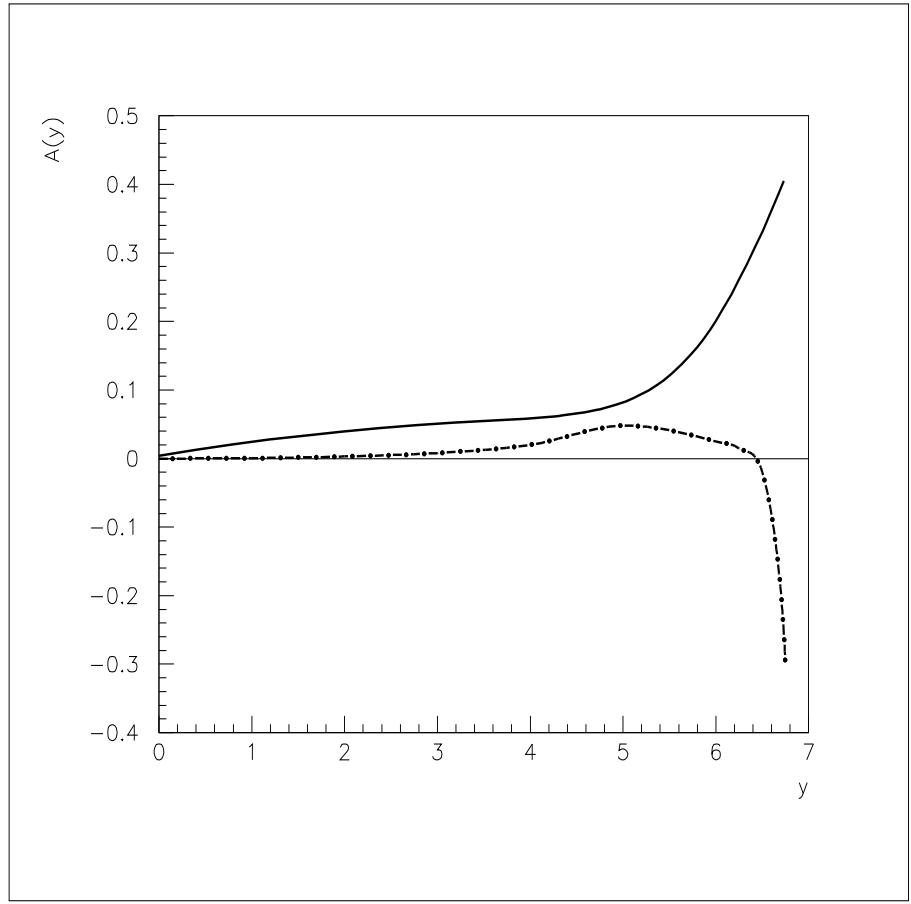,height=12cm,width=12cm}}
\caption{\label{asym}$B^0/\bar{B^0}$ asymmetries at LHC energy
given by PYTHIA \protect\cite{norrbin} and by QGS model; the value 
$\Delta_{eff}$=0.3 correspods to the production of the mass of B-meson.}
\end{figure}

The A(y) dependence in fragmentation region $x_F \rightarrow$ 1 contradicts 
with all similar asymmetry measurements for D-meson spectra \cite{wa89,
selex,wa92,e791}. The intersection of inclusive spectra of different type 
of B-mesons gives the asymmetry passing trough zero at some $y_0$, while 
the measured spectra of leading particles are always higher then nonleading 
particle spectra, so the asymmetry is positive value. In opposite to PYTHIA 
predictions, the asymmetry calculated in the framework of QGSM is rising 
function up to $x_F \rightarrow$ 1. This behavior is usually
peculiar for the string approach because of so called "beam drag" effect.
The valuable asymmetry in central region given in QGSM prediction 
\cite{asymmetry} is not enough small for not to be taken into account at 
CP violation measurements. It looks important to consider both these
predictions in details and to discuss the probability of nonzero asymmetry 
in the production spectra at LHC energy.

\section{Conclusions}

 We have compared two approaches for the understanding of the heavy 
flavored particle production: one of them is mostly perturbative  and 
another one is nonperturbative at all.  This  comparison shows that some 
different  suggestion has to be done for low transverse momenta 
distributions of B-mesons to put into agreement the both model predictions 
at LHC energy. The contradicting dependences for $B^0/\bar{B^0}$ asymmetry 
in the B meson production spectra might be important for  the
CP violation measurements.

\section{Acknowledgments}

I would like to express my gratitude to  A.B.Kaidalov, A.Kharchilava and
S.Baranov for numerous discussions.
This work was supported by DFG grant   $N^{\circ}$  436 RUS 113/332/O(R).

\end{document}